\documentclass[showpacs,twocolumn,pra,longbibliography]{revtex4-1}

\usepackage{epsf}
\usepackage{amsmath}
\usepackage{amsfonts}
\usepackage{amssymb}
\usepackage{bm}
\usepackage{bbm}
\usepackage{nicefrac}
\usepackage{cancel}
\usepackage{xcolor}
\usepackage{maybemath}
\usepackage{pifont}
\usepackage{afterpage}
\usepackage{subfloat}

\usepackage{graphicx}
\usepackage{epsfig}
\usepackage{latexsym}
\usepackage{dcolumn}
\usepackage{pifont}
\usepackage{tikz}

\definecolor{kellygreen}{rgb}{0.3, 0.73, 0.09}

\definecolor{garrosgreen}{rgb}{0.1, 0.4, 0.1}
\definecolor{dartmouthgreen}{rgb}{0.05, 0.5, 0.06}

\definecolor{duelferred}{rgb}{0.7, 0.2, 0.1}
\definecolor{cambridgeblue}{rgb}{0.1, 0.3, 1.0}
\definecolor{oxfordblue}{rgb}{0.05, 0.2, 0.7}

\def\vdw{van der Waals}

\def\HFS{{\mathrm{HFS}}}

\def\LS{{\mathrm{LS}}}

\def\vdW{{\mathrm{vdW}}}

\def\calH{{\mathcal H}}
\def\calL{{\mathcal L}}
\def\calF{{\mathcal F}}

\def\calV{{\mathcal V}}

\def\vecc{\vec}

\newcommand{\Ecr}{E_{\rm cr}}
\newcommand{\gcr}{g_{\rm cr}}

\makeatletter

\newcommand{\Rmnum}[1]{\expandafter\@slowromancap\romannumeral #1@}
\makeatother

\newcolumntype{.}{D{x}{}{-1}}

\begin{document}

\newcommand{\addrROLLA}{Department of Physics,
Missouri University of Science and Technology,
Rolla, Missouri 65409-0640, USA}

\title{Adjacency Graphs and Long-Range Interactions of Atoms in Quasi-Degenerate
States: Applied Graph Theory}

\author{C. M. Adhikari}
\affiliation{\addrROLLA}

\author{V. Debierre}
\affiliation{\addrROLLA}

\author{U. D. Jentschura}
\affiliation{\addrROLLA}

\begin{abstract}
We analyze, in general terms, the evolution of 
energy levels in quantum mechanics, as a function
of a coupling parameter, and demonstrate the 
possibility of level crossings in systems described 
by irreducible matrices. In long-range interactions, 
the coupling parameter is the interatomic distance.
We demonstrate the utility of adjacency matrices 
and adjacency graphs in the analysis of 
``hidden'' symmetries of a problem; these allow
us to break reducible matrices into 
irreducible subcomponents. 
A possible 
breakdown of the no-crossing theorem for 
higher-dimensional irreducible matrices is indicated,
and an application to the 
$2S$--$2S$ interaction in hydrogen is briefly described.
The analysis of interatomic interactions in this 
system is important for further progress on optical
measurements of the $2S$ hyperfine splitting.
\end{abstract}

\maketitle

%
%
\section{Introduction}
\label{sec1}

In quantum mechanical systems described by a ($2 \times 2$)-matrix, 
no level crossings can typically occur~\cite{CTDiLa1978vol1,CTDiLa1978vol2}. 
This is known as the ``no level crossing theorem'' 
and often illustrated on the basis of the 
simple $(2 \times 2)$-model Hamiltonian matrix 
\begin{equation}
H' = H + P =
\left( \begin{array}{cc} 
E_1 & 0 \\ 0 & E_2 
\end{array} \right) +
\left( \begin{array}{cc} 
0 & C \, g \\ C \, g & 0 
\end{array} \right)  \,,
\end{equation}
where $E_1$ and $E_2$ are the unperturbed energy levels,
$C$ is a parameter, and $g$ is the coupling constant.
The energy levels are 
\begin{equation} 
E_\pm = \frac12 \, (E_1 + E_2) +
\frac12 \, \sqrt{ (E_1 - E_2)^2 + 4 (C \, g)^2 }  \,.
\end{equation}
As a function of $g$, one obtains two hyperbolas,
with the distance of ``closest approach'' between the energy 
levels occurring for $g=0$, with a separation 
$| E_+ - E_-| = |E_1 - E_2|$.
For a level crossing to occur at $g=0$,
one has to have $E_1 = E_2$.
The larger the perturbation, the more the 
energy levels ``repel'' each other.

However, the situation is less clear for more complex 
systems involving more than two energy levels.
To this end, we shall analyze a higher-rank 
matrix which describes energy levels some of which 
repel each other on the basis of inter-level couplings,
in a system which obviously can be broken 
into smaller subcomponents (i.e., the 
Hamiltonian is a reducible matrix
having irreducible submatrices). As the 
levels in the irreducible 
subsystems evolve from the weak-coupling to the 
strong-coupling regime, those coming 
from different irreducible submatrices  cross.
When additional couplings are introduced between the 
subsystems, the matrix becomes irreducible.
In this case, we shall demonstrate
that some of the level crossings are avoided,
but not all. Our example will be based on a 
$(6 \times 6)$-matrix.

Another question which sometimes occurs in the 
analysis of interatomic interactions,
and other contexts in quantum mechanics,
concerns the reducibility of a matrix.
Reducible tensors are usually introduced in the 
context of the rotation group.
Under a rotation, scalars transform 
into scalars, vectors transform into vectors,
quadrupole tensors transform into
quadrupole tensors, and so on.
It means that a matrix representation 
of the rotation would have an obvious 
block structure when formulated in terms
of the irreducible tensor components.
For example, a trivially reducible matrix is 
\begin{equation}
H'' = 
\left( \begin{array}{ccc}
E_1 & C\, g & 0 \\[0.1133ex]
C\,g & E_2 & 0 \\[0.1133ex]
0 & 0 & E_3 
\end{array} \right)  \,,
\end{equation}
as it can obviously be broken into an
upper $(2 \times 2)$ submatrix equal to $H'$,
and a lower $(1 \times 1)$ submatrix 
just consisting of the uncoupled energy level $E_3$.

The question of whether a higher-dimensional
matrix is reducible, can be far less trivial to analyze.
For example, in a $(24 \times 24)$ matrix,
as has been recently encountered in our 
analysis of the $2S$--$2S$ hyperfine-resolved interactions
in hydrogen~\cite{JeEtAl2016vdWii}, entries can follow a rather 
irregular pattern, and the analysis then becomes far less
trivial.
The possibility to break up a matrix into irreducible 
subcomponents is equivalent to a search for 
``hidden'' symmetries of the interaction which 
imply that only sublevels of specific symmetry are coupled.

After a brief look at level crossings in Sec.~\ref{sec2},
we continue with an analysis of irreducible (sub-)matrices
in Sec.~\ref{sec3}. An application 
of the concepts developed  
to the $2S$--$2S$ hyperfine interaction in hydrogen 
is briefly described in Sec.~\ref{sec4}.

%
%
\section{Couplings and Level Crossings}
\label{sec2}

\begin{figure}[t!]
\begin{center}
\begin{minipage}{0.91\linewidth}
\begin{center}
\includegraphics[width=1.0\linewidth]{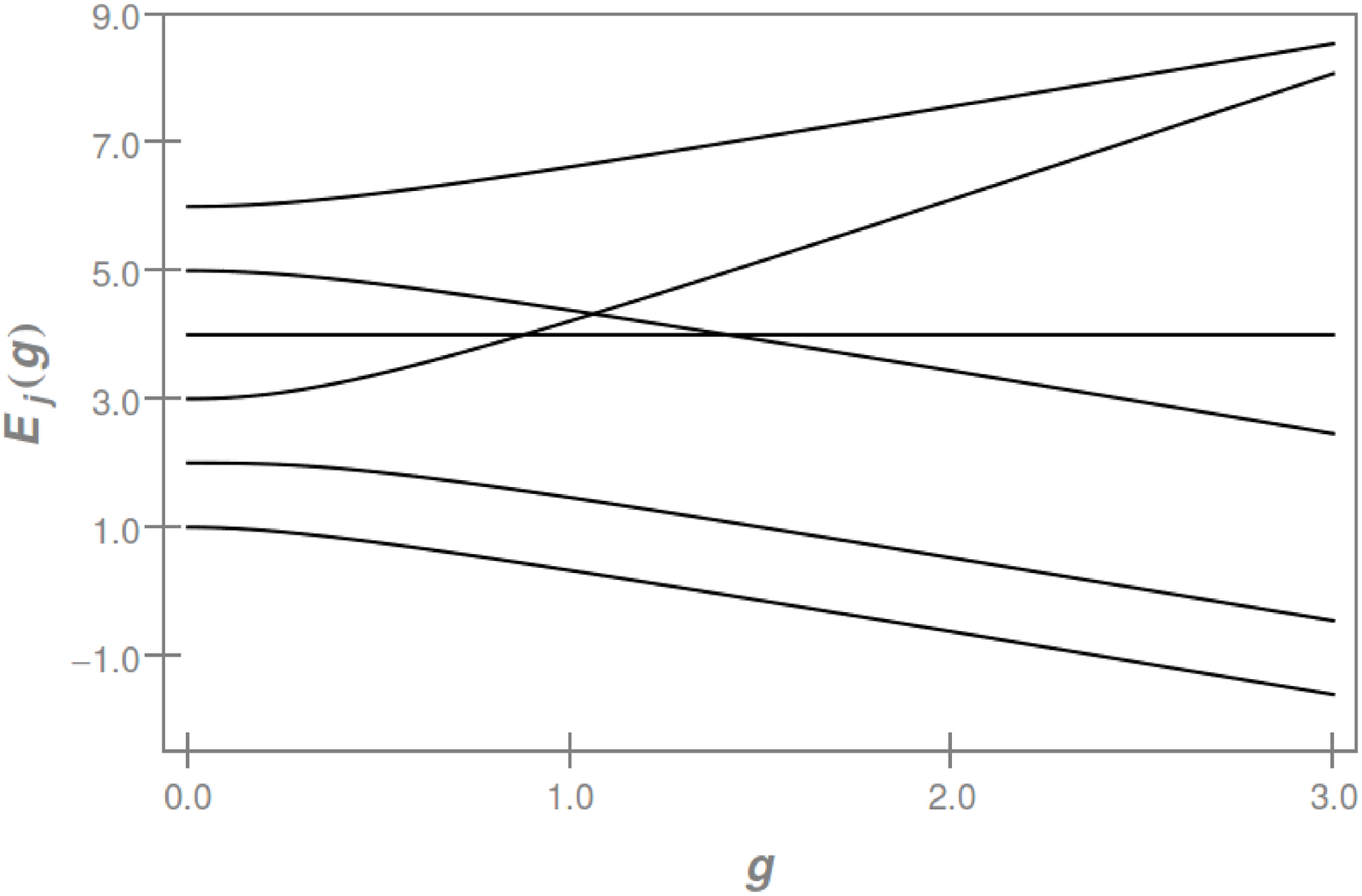}
\end{center}
\end{minipage}
\caption{\label{fig1} Evolution of the energy 
levels $E_j(g)$ of the matrix $H_0$ given in 
Eq.~\eqref{H0}, for the 
parameter choice given in Eq.~\eqref{choice}.
One can clearly discern the mutual ``repulsion''
between the lowest three energy levels
$E_{1,2,3}$, stemming from the upper $(3 \times 3)$-block 
of the matrix~\eqref{H0}, and the same repulsion among the 
highest energies $E_{4,5}$, 
stemming from the upper $(2 \times 2)$-block 
of the matrix~\eqref{H0}.
The level crossings occur with respect to the 
uncoupled level $E_3$, which is independent of $g$.}
\end{center}
\end{figure}

Let us consider the 6 x 6 matrix
\begin{equation}
\label{H0}
H_0 = 
\left( \begin{array}{cccccc}
E_1 & C_1 \, g & C_1 \, g & 0 & 0 & 0 \\
C_1 \, g & E_2 & C_1 \, g & 0 & 0 & 0 \\
C_1 \, g & C_1 \, g & E_3 & 0 & 0 & 0 \\
0 & 0 & 0 & E_4 & 0 & 0 \\
0 & 0 & 0 & 0 & E_5 & C_1 \, g \\
0 & 0 & 0 & 0 & C_1 \, g & E_6
\end{array} \right)
\end{equation}
This matrix consists of a mutually coupled 
(irreducible) upper 
$(3 \times 3)$-block, an irreducible 
lower $(2 \times 2)$-block, 
and one uncoupled state in the middle,
with energy $E_3$.
For the choice 
\begin{equation}
\label{choice}
E_j = j \,, \qquad C_1 = 1 \,,
\end{equation}
the evolution of the eigenenergies 
$E_j \to E_j(g)$ is analyzed in Fig.~\ref{fig1}.
Specifically, the level crossings occur at 
\begin{subequations}
\begin{align}
\label{crossing1}
E_3(g') =& \; E_4(g') = 4 \,, \qquad g' = 0.879\,385 \,,
\\[0.1133ex]
\label{crossing2}
E_3(g'') =& \; E_5(g'') = 4.326\,328  \,,
\nonumber\\[0.1133ex]
g'' =& \; 1.061\,840 \,,
\\[0.1133ex]
\label{crossing3}
E_4(g''') =& \; E_5(g''') = 4, \qquad g''' = \sqrt{2} \,.
\end{align}
\end{subequations}

\begin{figure}[t!]
\begin{center}
\begin{minipage}{0.91\linewidth}
\begin{center}
\includegraphics[width=1.0\linewidth]{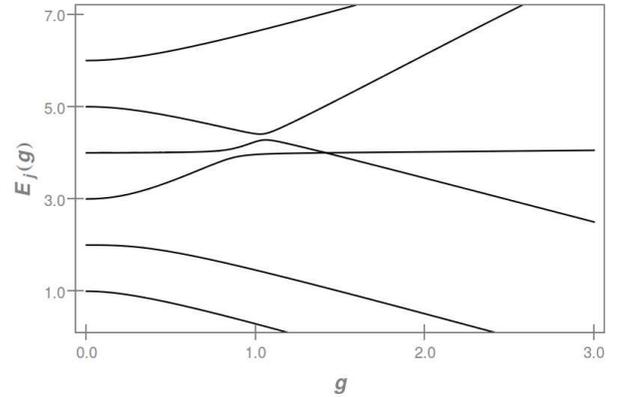}
\end{center}
\end{minipage}
\caption{\label{fig2} Evolution of the energy
levels $E_j(g)$ of the matrix $H$ given in
Eq.~\eqref{H}, for the parameter choices given in 
Eqs.~\eqref{choice} and~\eqref{choice2}.
In comparison to Fig.~\ref{fig1},
the ordinate axis is compressed in order to 
focus on the level crossings.
The crossings~\eqref{crossing1} and~\eqref{crossing2} 
have turned into anticrossings, in view of 
the mutual level repulsion as the inter-level 
couplings are introduced, in 
accordance with the no-crossing theorem.
However, the crossing~\eqref{crossing3} 
is retained (with a slightly different values of $g'''$),  
with the twist that it takes place between $E_3$ and $E_4$ 
this time (instead of $E_4$ and $E_5$ as in the previous case). 
This change is due to the fact that the crossing~\eqref{crossing1} 
between $E_3$ and $E_4$ and the crossing~\eqref{crossing2} between $E_3$ and $E_5$
are now avoided.}
\end{center}
\end{figure}

\begin{figure}[t!]
\begin{center}
\begin{minipage}{0.91\linewidth}
\begin{center}
\includegraphics[width=1.0\linewidth]{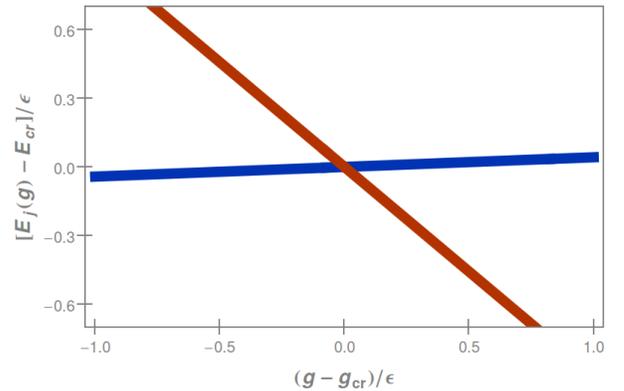}
\end{center}
\end{minipage}
\caption{\label{fig3} Close-up of Fig.~\ref{fig2} 
in the region $E_3(g) \approx E_4(g) \approx 4$,
and $g \approx g_{\rm cr} = \sqrt{2}$, with 
$\epsilon = 10^{-128}$. This plot was obtained 
using extended-precision arithmetic, using a 
computer algebra system~\cite{Wo1999}.
The observed numerical behavior is consistent
with the persistence of the level crossing 
for the irreducible matrix.}
\end{center}
\end{figure}

Let us now add a further perturbation $H_1$,
\begin{equation}
\label{H1}
H_1 =
\left( \begin{array}{cccccc}
0 & 0 & 0 & C_2 \, g & 0 & C_2 \, g  \\
0 & 0 & 0 & 0 & 0 & 0 \\
0 & 0 & 0 & 0 & 0 & 0 \\
C_2 \, g & 0 & 0 & 0 & 0 & 0 \\
0 & 0 & 0 & 0 & 0 & 0 \\
C_2 \, g & 0 & 0 & 0 & 0 & 0 
\end{array} \right) \,,
\end{equation}
where $C_2$ is another parameter,
to obtain the total Hamiltonian
\begin{equation}
\label{H}
H = H_0 + H_1 =
\left( \begin{array}{cccccc}
E_1 & C_1 \, g & C_1 \, g & C_2 \, g  & 0 & C_2 \, g \\
C_1 \, g & E_2 & C_1 \, g & 0 & 0 & 0 \\
C_1 \, g & C_1 \, g & E_3 & 0 & 0 & 0 \\
C_2 \, g & 0 & 0 & E_4 & 0 & 0 \\
0 & 0 & 0 & 0 & E_5 & C_1 \, g \\
C_2 \, g  & 0 & 0 & 0 & C_1 \, g & E_6
\end{array} \right) \,.
\end{equation}
In the total Hamiltonian $H$, the 
previously uncoupled level $E_4$ is now 
coupled to the upper $(3 \times 3)$ block
by the term $C_2$, and an additional
coupling between the lower $(2 \times 2)$ block 
and the upper $(3 \times 3)$ block
is introduced in the extreme upper right 
and lower left corners of the matrices $H_1$
and $H$. In Fig.~\ref{fig2}, we 
study the evolution of the energy levels of $H$
for the parameter choice
\begin{equation}
\label{choice2}
C_2 = \frac{3}{10} \,.
\end{equation}
It is clearly seen that the level crossings~\eqref{crossing1}
and~\eqref{crossing2} now turn into avoided crossings,
while the crossing~\eqref{crossing3} is retained, 
but now occurs between $E_3$ and $E_4$ and not between 
$E_3$ and $E_4$. This difference is due to the avoided crossings.

The value of the energy at the crossing occurs at
the coupling $g = \gcr$,
\begin{equation}
\Ecr = E_3(\gcr) = E_4(\gcr) = 4 \,,
\qquad
\gcr = \sqrt{2} \,.
\end{equation}
We have verified (see Fig.~\ref{fig3}) that the 
crossing persists under the use of extended-precision
arithmetic, where the parameter $\epsilon = 10^{-128}$ 
(on the level of Fortran ``hexadecuple precision'')
is employed in a numerical calculation of the 
eigenvalue near the crossing point, in order to 
ensure that the persistence of the crossing is 
not an artefact due to an insufficient numerical 
accuracy in the calculation. One might otherwise 
conjecture that the ``crossing'' would turn
into an ``avoided crossing'' when looking at the crossing 
point with finer numerical resolution.

For $C_1 = 1$, as a function of $C_2$, the eigenvectors 
at the degenerate eigenvalue $\Ecr = 4$ 
(where the crossing occurs) can be determined 
analytically; they read as
\begin{subequations}
\label{v3v4}
\begin{align}
v_3 =& \; \left( 0, 
-\frac{C_2}{1 + \sqrt{2}},
-\frac{\sqrt{2} \, C_2}{1 + \sqrt{2}},
0, -\sqrt{2}, 1 \right) \,,
\\[0.1133ex]
v_4 =& \; \left( 0, 
-\frac{C_2}{1 + \sqrt{2}},
-\frac{\sqrt{2} \, C_2}{1 + \sqrt{2}},
1, 0, 0 \right) \,.
\end{align}
\end{subequations}
A comparison of Fig.~\ref{fig1} to Fig.~\ref{fig2} reveals 
that the crossing  ``actually'' occurs between the levels 
$4$ and $5$. According to 
the adjacency graph in Fig.~\ref{fig5},
the levels $4$ and $5$ are the most distant 
ones in comparison to the 
levels $2$ and $3$ which constitute the 
$C_2$-dependent ``admixtures'' at the crossing.
According to Eq.~\eqref{v3v4},
Furthermore, in the limit $C_2 \to 0$, 
the eigenvectors $v_3$ and $v_4$ 
given in Eq.~\eqref{v3v4} have contributions 
only from the unperturbed levels $4$, $5$, and $6$;
the latter are not directly coupled to 
the levels $2$ and $3$ in the adjacency graph in Fig.~\ref{fig5}.
Apparently, the no-crossing theorem
discussed in~Ref.~\cite{wikiavoidedcross}
does not hold for higher-dimensional matrices,
while crossings in $2 \times 2$ matrices are strictly 
avoided in view of this theorem
(see Chap.~79 of Ref.~\cite{LaLi1958vol3}).

A comparison of Figs.~\ref{fig1} and~\ref{fig2} 
reveals that the number of crossings is seen to be reduced for the 
case of the irreducible Hamiltonian matrix,
but it is not zero.

%
%
\section{Finding Irreducible Submatrices}
\label{sec3}

We shall briefly discuss how to establish, by a formal,
generalizable, method, that the matrix given 
in Eq.~\eqref{H0} is reducible, while the 
matrix~\eqref{H} is irreducible.

Let us look at a general $(n \times n)$ matrix
and associate it with the flight plan
of a specific airline, with 
a nonvanishing entry, equal to unity,
at position $(i,j)$,
denoting the existence of a direct flight
between the cities $i$ and $j$.
If the matrix element $(i,j)$ is zero, then 
no such direct connection exists.
This matrix is known as the ``adjacency 
matrix'' $U$ of the airline connection.
A nonvanishing entry at position $(i,i)$ could be 
interpreted as a ``sightseeing flight'' starting and 
ending at city $i$.
There could be 
an indirect coupling between cities $i$ and $j$,
if not by a direct flight, then via a connection 
through some city $k$.
If there is a connection with one
intermediate stop, then it is obvious that the 
square of the adjacency matrix 
will have a unit entry at position $(i,j)$. 
Nonzero entries in $U^2$ represent the 
cities that connect with connecting flights
(one intermediate stop only). 
More specifically, the entries in the square of the 
adjacency matrix count the number of 
possibilities that one can fly from city $i$
to city $j$ with exactly one intermediate stop.
If the airline serves $n$ airports
and one cannot go from city $i$ to city $j$ 
with $n-1$ intermediate stops, then one 
cannot go city $i$ to city $j$ at all.
One has exhausted the possibilities.
Let $U$ denote the adjacency matrix.
It means that if the matrix 
\begin{equation}
\label{A}
A = \sum_{i=1}^n U^i = U + U^2 + \ldots + U^n 
\end{equation}
still has a zero entry at position $(i,j)$,
then the airline must be serving at least two disconnected
sets of destinations; this in turn is equivalent 
to showing that the adjacency matrix is reducible.
The algorithm for testing the 
reducibility of an input matrix $M$ is now clear.
One replaces all nonzero entries in the input 
matrix $M$ by unity,
obtaining the adjacency matrix $U$.
One then calculates the accumulated 
adjacency matrix $A$ according to Eq.~\eqref{A}.
If there are zero entries in $A$, then $M$ must 
be reducible.

\begin{figure}[t!]
\begin{center}
\begin{minipage}{0.91\linewidth}
\begin{center}
\includegraphics[width=1.0\linewidth]{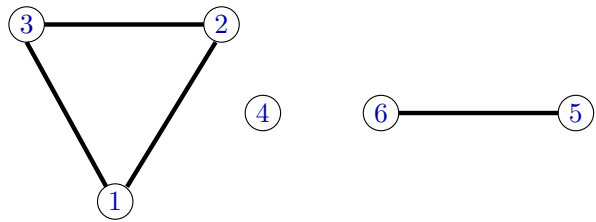}
\end{center}
\end{minipage}
\caption{\label{fig4}Adjacency graph for the 
matrix $U_0$ given in Eq.~\eqref{U0}.}
\end{center}
\end{figure}

\begin{figure}[t!]
\begin{center}
\begin{minipage}{0.91\linewidth}
\begin{center}
\includegraphics[width=1.0\linewidth]{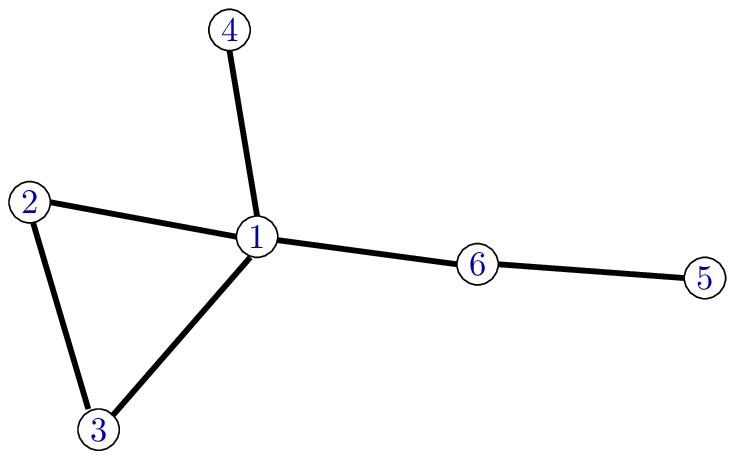}
\end{center}
\end{minipage}
\caption{\label{fig5} Adjacency graph for the
matrix $U$ given in~\eqref{matU}.}
\end{center}
\end{figure}

\begin{figure}[t!]
\begin{center}
\begin{minipage}{0.91\linewidth}
\begin{center}
\includegraphics[width=1.0\linewidth]{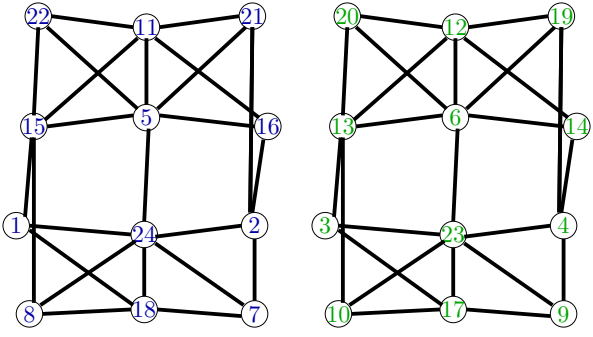}
\end{center}
\end{minipage}
\caption{\label{fig6} Adjacency graph for the
matrix $U_{F_z=0}$ given in Eq.~\eqref{UFz0}.}
\end{center}
\end{figure}

The adjacency matrix $U_0$ for $H_0$ given in Eq.~\eqref{H0} is
\begin{equation}
\label{U0}
U_0 = 
\left( \begin{array}{cccccc}
1 & 1 & 1 & 0 & 0 & 0 \\
1 & 1 & 1 & 0 & 0 & 0 \\
1 & 1 & 1 & 0 & 0 & 0 \\
0 & 0 & 0 & 1 & 0 & 0 \\
0 & 0 & 0 & 0 & 1 & 1 \\
0 & 0 & 0 & 0 & 1 & 1 
\end{array} \right) \,,
\end{equation}
resulting in 
\begin{equation}
A_0 = \sum_{i=1}^6 U_0^i = 
\left( \begin{array}{cccccc}
364 & 364 & 364 & 0 & 0 & 0 \\
364 & 364 & 364 & 0 & 0 & 0 \\
364 & 364 & 364 & 0 & 0 & 0 \\
0 & 0 & 0 & 6 & 0 & 0 \\
0 & 0 & 0 & 0 & 63 & 63 \\
0 & 0 & 0 & 0 & 63 & 63
\end{array} \right) \,,
\end{equation}
clearly displaying the reducibility and the 
three submatrices.
The corresponding adjacency graph is 
given in Fig.~\ref{fig4}.
These observations only confirm the 
intuitive understanding gathered by 
inspection of $H_0$.

For the matrix $H$ given in Eq.~\eqref{H},
the adjacency matrix is
\begin{equation}
\label{matU}
U =
\left( \begin{array}{cccccc}
1 & 1 & 1 & 1 & 0 & 1 \\
1 & 1 & 1 & 0 & 0 & 0 \\
1 & 1 & 1 & 0 & 0 & 0 \\
1 & 0 & 0 & 1 & 0 & 0 \\
0 & 0 & 0 & 0 & 1 & 1 \\
1 & 0 & 0 & 0 & 1 & 1
\end{array} \right) \,,
\end{equation}
resulting in
\begin{equation}
\label{matA}
A = \sum_{i=1}^6 U^i =
\left( \begin{array}{cccccc}
836 & 604 & 604 & 354 & 178 & 426 \\
604 & 453 & 453 & 250 & 106 & 284 \\
604 & 453 & 453 & 250 & 106 & 284 \\
354 & 250 & 250 & 158 &  72 & 178 \\
178 & 106 & 106 &  72 &  90 & 142 \\
426 & 284 & 284 & 178 & 142 & 268
\end{array} \right) \,,
\end{equation}
which is fully populated 
The corresponding adjacency graph is 
given in Fig.~\ref{fig5}.
The accumulated adjacency matrix $A$ is fully populated,
demonstrating the irreducibility of $H$.

%
%
\section{$\bm{2S}$--$\bm{2S}$ Interaction in Hydrogen}
\label{sec4}

The aim is to analyze the interaction of two excited 
hydrogen atoms in the metastable $2S$ state.
We note that the
$2S$--$2S$ \vdw{} interaction has been analyzed before in
Refs.~\cite{JoEtAl2002,SiKoHa2011}, but without any reference to the resolution
of the hyperfine splitting.
The Hamiltonian for the two-atom system is 
\begin{equation}
\label{HAB}
H = H_{\LS,A} + H_{\LS,B} +
H_{\HFS,A} + H_{\HFS,B} + H_{\vdW} \,.
\end{equation}
Here, $H_{\rm LS}$ is the Lamb shift Hamiltonian,
while $H_{\rm HFS}$ describes hyperfine effects;
these Hamiltonians have to be added for atoms $A$ and $B$.
In SI units, they are given as follows,
\begin{subequations}
\begin{align}
\label{HHFS}
H_{\rm HFS}=&\frac{\mu_0}{4\pi}\mu_B\mu_N\, g_s g_p
\sum_{i=A,B}\left[
\frac{8\pi}{3}\vecc{S}_i\cdot\vecc{I}_i\,\delta^3\left(\vecc{r}_i\right)\right.
\nonumber\\
& \left. + \frac{3\left(\vecc{S}_i\cdot\vecc{r}_i\right)
\left(\vecc{I}_i\cdot\vecc{r}_i\right)-\vecc{S}_i\cdot\vecc{I}_i\,\vecc{r}_i^2}%
{\left|\vecc{r}_i\right|^5}+\frac{\vecc{L}_i\cdot\vecc{I}_i}{\left|\vecc{r}_i\right|^3}\right] \,,
\\
\label{HLS}
H_{\rm LS}=&\frac{4}{3}\alpha^2\,m c^2\left(\frac{\hbar}{m c}\right)^3\ln\left(\alpha^{-2}\right)
\sum_{i=A,B}\delta^3\left(\vecc{r}_i\right) \,,
\\[0.133ex]
\label{vdw}
H_{\rm vdW} =& \;
\alpha \,\hbar c\, \frac{x_A \, x_B + y_A \, y_B - 2 \, z_A \, z_B}{R^3}  \,.
\end{align}
\end{subequations}
The symbols are explained as follows:
$\alpha$ is the fine-structure constant, $m $ denotes the electron mass.
The operators
$\vecc{r}_i$, $\vecc{p}_i$ and $\vecc{L}_i$ are the position (relative to
the respective nuclei), linear momentum and orbital angular momentum operators
for electron $i$, while $\vecc{S}_i$ is the spin operator for electron $i$
and $\vecc{I}_i$ is the spin operator for proton $i$ [both are
dimensionless]. Electronic and protonic $g$ factors are
$g_s\simeq2.002\,319$ and $g_p\simeq5.585\,695$, while
$\mu_B\simeq9.274\,010\,\times10^{-24}\,\mathrm{A m}^2$ is the Bohr magneton
and $\mu_N\simeq5.050\,784\,\times10^{-27}\,\mathrm{A m}^2$ is the nuclear
magneton.  Of course, the subscripts $A$ and $B$ refer to the relative coordinates within
the two atoms. $R$ is the interatomic distance. 
$H_{\rm LS}$ shifts $S$ states relative to $P$ states by the Lamb shift, which
is given in Eq.~\eqref{HLS} in 
the Welton approximation~\cite{ItZu1980}, which is convenient
within the formalism used for the evaluation of matrix elements.
The important property of $H_{\rm LS}$ is that it shifts
$S$ states upward in relation to $P$ states. The prefactor multiplying the
Dirac-$\delta$ can be adjusted to the observed Lamb shift
splitting. Indeed, for
the final calculation of energy shifts, one conveniently replaces
\begin{multline}
\label{defcalL}
\langle 2S_{1/2} | H_{LS} | 2S_{1/2} \rangle -
\langle 2P_{1/2} | H_{LS} | 2P_{1/2} \rangle \\
=\frac{4\alpha}{3 \pi}\, \frac{\alpha^4}{8} \, m \,c^2 \, \ln( \alpha^{-2})
\to \calL \,,
\end{multline}
where $\calL = h \times 1057.845(9) \, {\rm MHz}$ is the
``classic'' $2S$--$2P_{1/2}$ Lamb shift \cite{LuPi1981}
($m$ is the electron mass, $c$ is the speed of light,
and $h$ is Planck's constant).
In the Hamiltonian (\ref{HAB}) the origin of energies
is taken at the hyperfine center of the $2P_{1/2}$ levels.

The coupling scheme for the atomic levels entails 
that the orbital angular momentum $\vec L_i$ should be 
added to the electron spin to give 
the total angular momentum $\vec J_i$, 
then $\vec J_i$ is added to the nuclear spin $\vec I_i$ to give $\vec F_i$.
This vector coupling has to be done for both atoms $i = A, B$,
and then $\vec F =  \vec F_A + \vec F_B$
(orbital$+$spin$+$nuclear angular momentum,
summed over both atoms $A$ and $B$). 
One can show relatively easily that the 
$z$ component $F_z$ of the total angular momentum 
is conserved, i.e., $F_z$ commutes with the Hamiltonian.

We restrict the discussion to states with 
total angular momentum $J = 1/2$,
i.e., to the $2S$ and $2P_{1/2}$ states which are 
displaced from each other only by the Lamb shift.
States displaced by the fine structure are 
subdominant because $\calF \gg \cal L$
where $\calF = \alpha^4 m c^2/32$ is the 
$2P$ fine-structure interval.

Let each atom be in a state $| \ell_i, F_i, F_{z,i} \rangle$,
with $i= A,B$ (here, $\ell_i$ is the orbital angular momentum
quantum number). The two-atom system occupies the states
$| (\ell_A, F_A, F_{z,A})_A \,
(\ell_B, F_B, F_{z,B})_B \rangle$.
We have four $S$ states ($F=0$ and $F=1$),
and four $P$ states ($F=0$ and $F=1$),
for each atom, making for a total of eight states.
For two atoms, one thus has 64 states in the 
($n=2$)--($n=2$) manifold with $J = 1/2$.

Now, since $F_z = F_{z,A} + F_{z,B}$ is a conserved quantity,
we should classify states according to $F_z = \pm 2$,
$F_z = \pm 1$, and $F_z = 0$.
There are 4 states in the $F_z = \pm 2$ manifolds,
$16$ states in the  $F_z = \pm 1$ manifolds,
and a total of $24$ states in $F_z = 0$,
adding up to a total of 
$64 = 24 + 2 \times 16 + 2 \times 4$.
For $F_z = 0$, the matrix with $24 \times 24 = 576$ entries is hard to 
analyze. The question is whether or not one can find 
an additional symmetry that simplifies the analysis.
Such an additional symmetry would naturally lead to a
separation of the Hamiltonian into further irreducible 
submatrices, thus reducing the complexity of the 
computational task drastically.
It is precisely at this point that the methods 
discussed in Sec.~\ref{sec2} become useful.

To this end, we first 
order the states in the $F_z = 0$ manifold
according to increasing quantum numbers.
The state where atom $A$ is in an $S$ state
with $F_A = 0$, are given by
\begin{align}
| \Psi_{1} \rangle =& \; | (0,0,0)_A \, (0,0,0)_B \rangle \,, 
\nonumber\\[0.1133ex]
| \Psi_{2} \rangle =& \; | (0,0,0)_A \, (0,1,0)_B \rangle \,,
\nonumber\\[0.1133ex]
| \Psi_{3} \rangle =& \; | (0,0,0)_A \, (1,0,0)_B \rangle \,, \;
\nonumber\\[0.1133ex]
| \Psi_{4} \rangle =& \; | (0,0,0)_A \, (1,1,0)_B \rangle \,.
\end{align}
With atom $A$ in an $S$ state with $F_A = 1$, we have
\begin{align}
| \Psi_{5} \rangle =& \; | (0,1,-1)_A \, (0,1,1)_B \rangle \,, \;
\nonumber\\[0.1133ex]
| \Psi_{6} \rangle =& \; | (0,1,-1)_A \, (1,1,1)_B \rangle \,,
\nonumber\\[0.1133ex]
| \Psi_{7} \rangle =& \; | (0,1,0)_A \, (0,0,0)_B \rangle \,, \;
\nonumber\\[0.1133ex]
| \Psi_{8} \rangle =& \; | (0,1,0)_A \, (0,1,0)_B \rangle \,, \;
\nonumber\\[0.1133ex]
| \Psi_{9} \rangle =& \; | (0,1,0)_A \, (1,0,0)_B \rangle \,, \;
\nonumber\\[0.1133ex]
| \Psi_{10} \rangle =& \; | (0,1,0)_A \, (1,1,0)_B \rangle \,,
\nonumber\\[0.1133ex]
| \Psi_{11} \rangle =& \; | (0,1,1)_A \, (0,1,-1)_B \rangle \,,
\nonumber\\[0.1133ex]
| \Psi_{12} \rangle =& \; | (0,1,1)_A \, (1,1,-1)_B \rangle \,.
\end{align}
The states with atom $A$ in a $P_{1/2}$ state
(hyperfine singlet) are given as follows,
\begin{align}
| \Psi_{13} \rangle =& \; | (1,0,0)_A \, (0,0,0)_B \rangle \,, 
\nonumber\\[0.1133ex]
| \Psi_{14} \rangle =&\; | (1,0,0)_A \, (0,1,0)_B \rangle \,,
\nonumber\\[0.1133ex]
| \Psi_{15} \rangle =& \; | (1,0,0)_A \, (1,0,0)_B \rangle \,, 
\nonumber\\[0.1133ex]
| \Psi_{16} \rangle =& \; | (1,0,0)_A \, (1,1,0)_B \rangle \,.
\end{align} 
The states with atom $A$ in a $2P_{1/2}$ hyperfine triplet, 
are given by 
\begin{align}
| \Psi_{17} \rangle =& \; | (1,1,-1)_A \, (0,1,1)_B \rangle \,, 
\nonumber\\[0.1133ex]
| \Psi_{18} \rangle =& \; | (1,1,-1)_A \, (1,1,1)_B \rangle \,,
\nonumber\\[0.1133ex]
| \Psi_{19} \rangle =& \; | (1,1,0)_A \, (0,0,0)_B \rangle \,, 
\nonumber\\[0.1133ex]
| \Psi_{20} \rangle =& \; | (1,1,0)_A \, (0,1,0)_B \rangle \,,
\nonumber\\[0.1133ex]
| \Psi_{21} \rangle =& \; | (1,1,0)_A \, (1,0,0)_B \rangle \,,
\nonumber\\[0.1133ex]
| \Psi_{22} \rangle =& \; | (1,1,0)_A \, (1,1,0)_B \rangle \,,
\nonumber\\[0.1133ex]
| \Psi_{23} \rangle =&\; | (1,1,1)_A \, (0,1,-1)_B \rangle \,,
\nonumber\\[0.1133ex]
| \Psi_{24} \rangle =& \; | (1,1,1)_A \, (1,1,-1)_B \rangle \,.
\end{align}

\begin{figure*}[t!]
\begin{center}
\begin{minipage}{0.91\linewidth}
\begin{center}
\includegraphics[width=1.0\linewidth]{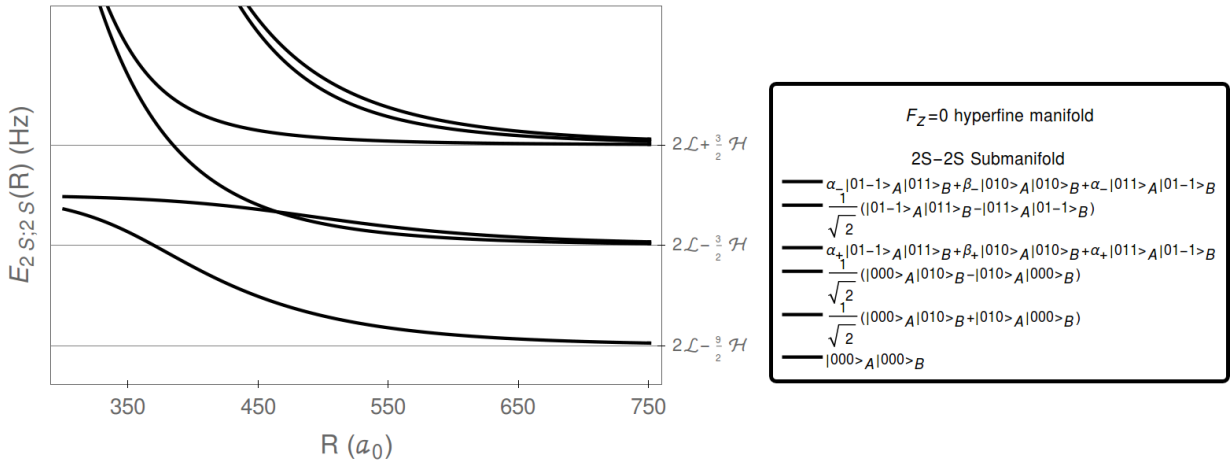}
\end{center}
\end{minipage}
\caption{\label{fig7} Energy levels of the $2S$--$2S$
states within the $F_z=0$ hyperfine manifold as a function of 
atomic separation $R$ (given in units of the Bohr radius $a_0$). 
The eigenstates in the legend are those relevant to 
the $R \to \infty$ asymptotic limit; 
for finite separation these states mix. There is 
one remaining level crossing even if the Hamiltonian 
matrix is irreducible. The coefficients 
$\alpha_\pm$ and $\beta_\pm$ are determined from
second-order perturbation theory and given by Eq.~(\ref{eq:AlphaBeta}). 
The states are labeled from top to bottom in the legend,
in the same order as they are relevant to the 
long-range asymptotics.}
\end{center}
\end{figure*}

The adjacency matrix $U_{F_z = 0}$ of the Hamiltonian~\eqref{HAB} in the 
$F_Z = 0$ manifold is equal to 
\tiny 
\begin{equation}
\label{UFz0}
\left(
\begin{array}{cccccccccccccccccccccccc}
 1 & 0 & 0 & 0 & 0 & 0 & 0 & 0 & 0 & 0 & 0 & 0 & 0 & 0 & 0 & 0 & 0 & 1 & 0 & 0 & 0 & 1 & 0 & 1 \\
 0 & 1 & 0 & 0 & 0 & 0 & 0 & 0 & 0 & 0 & 0 & 0 & 0 & 0 & 0 & 0 & 0 & 1 & 0 & 0 & 1 & 0 & 0 & 1 \\
 0 & 0 & 1 & 0 & 0 & 0 & 0 & 0 & 0 & 0 & 0 & 0 & 0 & 0 & 0 & 0 & 1 & 0 & 0 & 1 & 0 & 0 & 1 & 0 \\
 0 & 0 & 0 & 1 & 0 & 0 & 0 & 0 & 0 & 0 & 0 & 0 & 0 & 0 & 0 & 0 & 1 & 0 & 1 & 0 & 0 & 0 & 1 & 0 \\
 0 & 0 & 0 & 0 & 1 & 0 & 0 & 0 & 0 & 0 & 0 & 0 & 0 & 0 & 1 & 1 & 0 & 1 & 0 & 0 & 1 & 1 & 0 & 0 \\
 0 & 0 & 0 & 0 & 0 & 1 & 0 & 0 & 0 & 0 & 0 & 0 & 1 & 1 & 0 & 0 & 1 & 0 & 1 & 1 & 0 & 0 & 0 & 0 \\
 0 & 0 & 0 & 0 & 0 & 0 & 1 & 0 & 0 & 0 & 0 & 0 & 0 & 0 & 0 & 1 & 0 & 1 & 0 & 0 & 0 & 0 & 0 & 1 \\
 0 & 0 & 0 & 0 & 0 & 0 & 0 & 1 & 0 & 0 & 0 & 0 & 0 & 0 & 1 & 0 & 0 & 1 & 0 & 0 & 0 & 0 & 0 & 1 \\
 0 & 0 & 0 & 0 & 0 & 0 & 0 & 0 & 1 & 0 & 0 & 0 & 0 & 1 & 0 & 0 & 1 & 0 & 0 & 0 & 0 & 0 & 1 & 0 \\
 0 & 0 & 0 & 0 & 0 & 0 & 0 & 0 & 0 & 1 & 0 & 0 & 1 & 0 & 0 & 0 & 1 & 0 & 0 & 0 & 0 & 0 & 1 & 0 \\
 0 & 0 & 0 & 0 & 0 & 0 & 0 & 0 & 0 & 0 & 1 & 0 & 0 & 0 & 1 & 1 & 0 & 0 & 0 & 0 & 1 & 1 & 0 & 1 \\
 0 & 0 & 0 & 0 & 0 & 0 & 0 & 0 & 0 & 0 & 0 & 1 & 1 & 1 & 0 & 0 & 0 & 0 & 1 & 1 & 0 & 0 & 1 & 0 \\
 0 & 0 & 0 & 0 & 0 & 1 & 0 & 0 & 0 & 1 & 0 & 1 & 1 & 0 & 0 & 0 & 0 & 0 & 0 & 0 & 0 & 0 & 0 & 0 \\
 0 & 0 & 0 & 0 & 0 & 1 & 0 & 0 & 1 & 0 & 0 & 1 & 0 & 1 & 0 & 0 & 0 & 0 & 0 & 0 & 0 & 0 & 0 & 0 \\
 0 & 0 & 0 & 0 & 1 & 0 & 0 & 1 & 0 & 0 & 1 & 0 & 0 & 0 & 1 & 0 & 0 & 0 & 0 & 0 & 0 & 0 & 0 & 0 \\
 0 & 0 & 0 & 0 & 1 & 0 & 1 & 0 & 0 & 0 & 1 & 0 & 0 & 0 & 0 & 1 & 0 & 0 & 0 & 0 & 0 & 0 & 0 & 0 \\
 0 & 0 & 1 & 1 & 0 & 1 & 0 & 0 & 1 & 1 & 0 & 0 & 0 & 0 & 0 & 0 & 1 & 0 & 0 & 0 & 0 & 0 & 0 & 0 \\
 1 & 1 & 0 & 0 & 1 & 0 & 1 & 1 & 0 & 0 & 0 & 0 & 0 & 0 & 0 & 0 & 0 & 1 & 0 & 0 & 0 & 0 & 0 & 0 \\
 0 & 0 & 0 & 1 & 0 & 1 & 0 & 0 & 0 & 0 & 0 & 1 & 0 & 0 & 0 & 0 & 0 & 0 & 1 & 0 & 0 & 0 & 0 & 0 \\
 0 & 0 & 1 & 0 & 0 & 1 & 0 & 0 & 0 & 0 & 0 & 1 & 0 & 0 & 0 & 0 & 0 & 0 & 0 & 1 & 0 & 0 & 0 & 0 \\
 0 & 1 & 0 & 0 & 1 & 0 & 0 & 0 & 0 & 0 & 1 & 0 & 0 & 0 & 0 & 0 & 0 & 0 & 0 & 0 & 1 & 0 & 0 & 0 \\
 1 & 0 & 0 & 0 & 1 & 0 & 0 & 0 & 0 & 0 & 1 & 0 & 0 & 0 & 0 & 0 & 0 & 0 & 0 & 0 & 0 & 1 & 0 & 0 \\
 0 & 0 & 1 & 1 & 0 & 0 & 0 & 0 & 1 & 1 & 0 & 1 & 0 & 0 & 0 & 0 & 0 & 0 & 0 & 0 & 0 & 0 & 1 & 0 \\
 1 & 1 & 0 & 0 & 0 & 0 & 1 & 1 & 0 & 0 & 1 & 0 & 0 & 0 & 0 & 0 & 0 & 0 & 0 & 0 & 0 & 0 & 0 & 1 \\
\end{array}
\right) \,.
\end{equation}
\normalsize
The accumulated adjacency matrix has the structure
\begin{widetext}
\begin{equation}
\label{AFz0}
A_{F_z = 0} = \sum_{i=1}^{24} U^i \simeq \left(
\begin{array}{cccccccccccccccccccccccc}
 \chi & \chi & 0 & 0 & \chi & 0 & \chi & \chi & 0 & 0 & \chi & 0 & 0 & 0 & \chi & \chi & 0 & \chi & 0 & 0 & \chi & \chi & 0 & \chi \\
 \chi & \chi & 0 & 0 & \chi & 0 & \chi & \chi & 0 & 0 & \chi & 0 & 0 & 0 & \chi & \chi & 0 & \chi & 0 & 0 & \chi & \chi & 0 & \chi \\
 0 & 0 & \chi & \chi & 0 & \chi & 0 & 0 & \chi & \chi & 0 & \chi & \chi & \chi & 0 & 0 & \chi & 0 & \chi & \chi & 0 & 0 & \chi & 0 \\
 0 & 0 & \chi & \chi & 0 & \chi & 0 & 0 & \chi & \chi & 0 & \chi & \chi & \chi & 0 & 0 & \chi & 0 & \chi & \chi & 0 & 0 & \chi & 0 \\
 \chi & \chi & 0 & 0 & \chi & 0 & \chi & \chi & 0 & 0 & \chi & 0 & 0 & 0 & \chi & \chi & 0 & \chi & 0 & 0 & \chi & \chi & 0 & \chi \\
 0 & 0 & \chi & \chi & 0 & \chi & 0 & 0 & \chi & \chi & 0 & \chi & \chi & \chi & 0 & 0 & \chi & 0 & \chi & \chi & 0 & 0 & \chi & 0 \\
 \chi & \chi & 0 & 0 & \chi & 0 & \chi & \chi & 0 & 0 & \chi & 0 & 0 & 0 & \chi & \chi & 0 & \chi & 0 & 0 & \chi & \chi & 0 & \chi \\
 \chi & \chi & 0 & 0 & \chi & 0 & \chi & \chi & 0 & 0 & \chi & 0 & 0 & 0 & \chi & \chi & 0 & \chi & 0 & 0 & \chi & \chi & 0 & \chi \\
 0 & 0 & \chi & \chi & 0 & \chi & 0 & 0 & \chi & \chi & 0 & \chi & \chi & \chi & 0 & 0 & \chi & 0 & \chi & \chi & 0 & 0 & \chi & 0 \\
 0 & 0 & \chi & \chi & 0 & \chi & 0 & 0 & \chi & \chi & 0 & \chi & \chi & \chi & 0 & 0 & \chi & 0 & \chi & \chi & 0 & 0 & \chi & 0 \\
 \chi & \chi & 0 & 0 & \chi & 0 & \chi & \chi & 0 & 0 & \chi & 0 & 0 & 0 & \chi & \chi & 0 & \chi & 0 & 0 & \chi & \chi & 0 & \chi \\
 0 & 0 & \chi & \chi & 0 & \chi & 0 & 0 & \chi & \chi & 0 & \chi & \chi & \chi & 0 & 0 & \chi & 0 & \chi & \chi & 0 & 0 & \chi & 0 \\
 0 & 0 & \chi & \chi & 0 & \chi & 0 & 0 & \chi & \chi & 0 & \chi & \chi & \chi & 0 & 0 & \chi & 0 & \chi & \chi & 0 & 0 & \chi & 0 \\
 0 & 0 & \chi & \chi & 0 & \chi & 0 & 0 & \chi & \chi & 0 & \chi & \chi & \chi & 0 & 0 & \chi & 0 & \chi & \chi & 0 & 0 & \chi & 0 \\
 \chi & \chi & 0 & 0 & \chi & 0 & \chi & \chi & 0 & 0 & \chi & 0 & 0 & 0 & \chi & \chi & 0 & \chi & 0 & 0 & \chi & \chi & 0 & \chi \\
 \chi & \chi & 0 & 0 & \chi & 0 & \chi & \chi & 0 & 0 & \chi & 0 & 0 & 0 & \chi & \chi & 0 & \chi & 0 & 0 & \chi & \chi & 0 & \chi \\
 0 & 0 & \chi & \chi & 0 & \chi & 0 & 0 & \chi & \chi & 0 & \chi & \chi & \chi & 0 & 0 & \chi & 0 & \chi & \chi & 0 & 0 & \chi & 0 \\
 \chi & \chi & 0 & 0 & \chi & 0 & \chi & \chi & 0 & 0 & \chi & 0 & 0 & 0 & \chi & \chi & 0 & \chi & 0 & 0 & \chi & \chi & 0 & \chi \\
 0 & 0 & \chi & \chi & 0 & \chi & 0 & 0 & \chi & \chi & 0 & \chi & \chi & \chi & 0 & 0 & \chi & 0 & \chi & \chi & 0 & 0 & \chi & 0 \\
 0 & 0 & \chi & \chi & 0 & \chi & 0 & 0 & \chi & \chi & 0 & \chi & \chi & \chi & 0 & 0 & \chi & 0 & \chi & \chi & 0 & 0 & \chi & 0 \\
 \chi & \chi & 0 & 0 & \chi & 0 & \chi & \chi & 0 & 0 & \chi & 0 & 0 & 0 & \chi & \chi & 0 & \chi & 0 & 0 & \chi & \chi & 0 & \chi \\
 \chi & \chi & 0 & 0 & \chi & 0 & \chi & \chi & 0 & 0 & \chi & 0 & 0 & 0 & \chi & \chi & 0 & \chi & 0 & 0 & \chi & \chi & 0 & \chi \\
 0 & 0 & \chi & \chi & 0 & \chi & 0 & 0 & \chi & \chi & 0 & \chi & \chi & \chi & 0 & 0 & \chi & 0 & \chi & \chi & 0 & 0 & \chi & 0 \\
 \chi & \chi & 0 & 0 & \chi & 0 & \chi & \chi & 0 & 0 & \chi & 0 & 0 & 0 & \chi & \chi & 0 & \chi & 0 & 0 & \chi & \chi & 0 & \chi \\
\end{array}
\right) \,,
\end{equation}
\normalsize
\end{widetext}
where $\chi$ stands for any entry different from zero
(the $\chi$s are not all equal).
The adjacency graph given in Fig.~\ref{fig6}
confirms the presence of two irreducible submatrices
of $H_{F_z = 0}$.
Indeed, the two uncoupled subspaces are spanned by the 
states $| \Psi_{i} \rangle$
with $i = 1, 2, 5, 7, 8, 11, 15, 16, 18, 21, 22, 24$ (subspace~I),
and $| \Psi_{j} \rangle$
with $j = 3, 4, 6, 9, 10, 12, 13, 14, 17, 19, 20, 23$ (subspace~II).
An ordering of the eigenvalues reveals that 
one can have coupling among the $S$--$S$ 
and $P$--$P$ states (distributed among atoms $A$ and $B$), 
forming submanifold I,
and among all $S$--$P$ and $P$--$S$ states
(distributed among atoms $A$ and $B$), 
forming submanifold II.
In retrospect, the separation is perhaps clear, 
but it is less obvious at first glance.

It is then possible to redefine the 
levels from which the $12 \times 12$ Hamiltonian 
matrix is constructed, in the first submanifold
of $F_z = 0$ (the $S$--$S$ coupled states). Specifically, one defines
$| \Psi_{1}^{\left(\mathrm{\Rmnum{1}}\right)} \rangle = | \Psi_1 \rangle$,
$| \Psi_{2}^{\left(\mathrm{\Rmnum{1}}\right)} \rangle = | \Psi_2 \rangle$,
$| \Psi_{3}^{\left(\mathrm{\Rmnum{1}}\right)} \rangle = | \Psi_5 \rangle$,
$| \Psi_{4}^{\left(\mathrm{\Rmnum{1}}\right)} \rangle = | \Psi_7 \rangle$,
$| \Psi_{5}^{\left(\mathrm{\Rmnum{1}}\right)} \rangle = | \Psi_8 \rangle$,
$| \Psi_{6}^{\left(\mathrm{\Rmnum{1}}\right)} \rangle = | \Psi_{11} \rangle$,
$| \Psi_{7}^{\left(\mathrm{\Rmnum{1}}\right)} \rangle = | \Psi_{15} \rangle$,
$| \Psi_{8}^{\left(\mathrm{\Rmnum{1}}\right)} \rangle = | \Psi_{16} \rangle$,
$| \Psi_{9}^{\left(\mathrm{\Rmnum{1}}\right)} \rangle = | \Psi_{18} \rangle$,
$| \Psi_{10}^{\left(\mathrm{\Rmnum{1}}\right)} \rangle = | \Psi_{21} \rangle$,
$| \Psi_{11}^{\left(\mathrm{\Rmnum{1}}\right)} \rangle = | \Psi_{22} \rangle$,
and
$| \Psi_{12}^{\left(\mathrm{\Rmnum{1}}\right)} \rangle = | \Psi_{24} \rangle$.
Within the space spanned by the 
$| \Psi_{i}^{\left(\mathrm{\Rmnum{1}}\right)} \rangle$ with 
$i = 1,2, \dots, 12$, the Hamiltonian matrix has the structure
\begin{widetext}
\begin{align}
\label{mat}
H_{F_z = 0}^{\left(\mathrm{\Rmnum{1}}\right)} =& \; \left(
\begin{array}{cccccccccccc}
 2 \calL-\tfrac92 \calH & 0 & 0 & 0 & 0 & 0 & 0 & 0 & - \calV & 0 & -2 \calV & - \calV \\
 0 & 2 \calL-\tfrac32 \calH & 0 & 0 & 0 & 0 & 0 & 0 & \calV & -2 \calV & 0 & - \calV \\
 0 & 0 & 2 \calL+\tfrac32 \calH & 0 & 0 & 0 & - \calV & \calV & 2 \calV & - \calV & \calV & 0 \\
 0 & 0 & 0 & 2 \calL-\tfrac32 \calH & 0 & 0 & 0 & -2 \calV & - \calV & 0 & 0 & \calV \\
 0 & 0 & 0 & 0 & 2 \calL+\tfrac32 \calH & 0 & -2 \calV & 0 & \calV & 0 & 0 & \calV \\
 0 & 0 & 0 & 0 & 0 & 2 \calL+\tfrac32 \calH & - \calV & - \calV & 0 & \calV & \calV & 2 \calV\\
 0 & 0 & - \calV & 0 & -2 \calV & - \calV & -\tfrac32 \calH & 0 & 0 & 0 & 0 & 0 \\
 0 & 0 & \calV & -2 \calV & 0 & - \calV & 0 & -\tfrac12 \calH & 0 & 0 & 0 & 0 \\
 - \calV & \calV & 2 \calV & - \calV & \calV & 0 & 0 & 0 & \tfrac12 \calH & 0 & 0 & 0 \\
 0 & -2 \calV & - \calV & 0 & 0 & \calV & 0 & 0 & 0 & -\tfrac12 \calH & 0 & 0 \\
 -2 \calV & 0 & \calV & 0 & 0 & \calV & 0 & 0 & 0 & 0 & \tfrac12 \calH & 0 \\
 - \calV & - \calV & 0 & \calV & \calV & 2 \calV & 0 & 0 & 0 & 0 & 0 & \tfrac12 \calH
\end{array}
\right) \,,
\end{align}
\end{widetext}
where
\begin{equation}
\mathcal{V} = 3\,\alpha\,\hbar c\,\frac{a_0^2}{R^3}\
\end{equation}
is a parameter that describes the strength of the
\vdw{} interaction. Furthermore,
\begin{equation}
\calH = \frac{\alpha^4}{18} \, g_N \, \frac{m }{m_p}\,m \,c^2 
\end{equation}
with $\calH \approx h \times  59.1856114(22) \, {\rm MHz}$,
parameterizes the hyperfine splitting
($m_p$ is the proton mass).
A close-up of the six energetically highest, distance-dependent 
$S$--$S$ state energy levels, coupled through virtual $P$--$P$
states, is given in Fig.~\ref{fig7}
(Born-Oppenheimer potential energy curves). 
We have verified that the crossing between the 
second and third level (counted in ascending order 
of the unperturbed energy for $R \to \infty$) 
persists under a drastic increase of the 
numerical accuracy, much like for our model problem
(Fig.~\ref{fig3}). 
The coefficients used in the legend for this figure
are given by
\begin{subequations} \label{eq:AlphaBeta}
\begin{align}
\alpha_\pm&=2\sqrt{\frac{2}{33\pm\sqrt{33}}}\,,\\
\beta_\pm&=\mp\frac{\sqrt{33}\pm1}{\sqrt{2\left(33\pm\sqrt{33}\right)}}\,.
\end{align}
\end{subequations}
Despite the fact that subspace~I is irreducible, one observes
one level crossing, much in line with the discussion presented
in Sec.~\ref{sec2}.
Finer details of the calculation will be presented 
in an upcoming work \cite{JeEtAl2016vdWii}.
Specifically, for large $R$, we can point out that 
all of the level shifts of the 
states in Fig.~\ref{fig7} are found to be of order 
$\mathcal{V}^2/\mathcal{L}$ and are thus of second
order in $\calV$, proportional to $1/R^6$ but 
drastically enhanced in their numerical magnitude 
as compared to ``normal'' \vdw{} shifts 
due to the $1/\calL$ denominator.
For the $1S$--$1S$ interaction, the well-known 
result involves a shift of order
$\mathcal{V}^2/E_h$,
where $E_h$ is the Hartree energy.
In the limit of large $R$, the $2S$--$2S$ interaction 
is seen to be larger
by a factor $1/\alpha^3 \sim 10^6$, in view of the 
smaller energy denominator which only involves the 
Lamb shift.

%
%
\section{Conclusions}

Often, in physics, we need to resort to mathematical 
sophistications in order to uncover properties of a physical system 
hidden from us at first glance.
In our case, we find that 
adjacency matrices and adjacency 
graphs help determine the reducibility of a 
matrix, and, in the analysis of the hyperfine-resolved 
$2S$--$2S$ interaction, 
help determine the irreducible subspaces
into which we may break the total Hamiltonian.
We were able to identify an additional 
selection rule, which is 
relatively obvious {\em a posteriori},
namely, that couplings occur between 
$S$--$S$ and $P$--$P$ levels, and between 
$S$--$P$ and $P$--$S$ levels, but there are
no coupling joining the two submanifolds
(see Sec.~\ref{sec4}).
The size of the matrix is reduced from $24 \times 24$
to $12 \times 12$.
It is somewhat surprising that the seemingly 
easy problem of identifying the irreducible
submatrices of a Hamiltonian, 
involves a rather sophisticated concept 
like an adjacency matrix.

Our model problem, studied in Sec.~\ref{sec2} and~\ref{sec3},
reveals that level crossings can occur even in 
well-behaved quantum mechanical systems,
described by inter-level couplings
varying with some parameter.
For the long-range interaction 
between atoms, the inverse interatomic distance 
$1/R$ is such a coupling parameter.
In Fig.~\ref{fig2} (model problem),
and in Fig.~\ref{fig7} ($2S$--$2S$ states within the 
$F_z = 0$ submanifold of the hydrogen long-range interaction),
level crossings are clearly visible even if the 
Hamiltonian matrix is irreducible.
Our extended-precision numerical calculations 
(Fig.~\ref{fig3}) and the analytic structure of the
``crossing'' eigenvectors in Eq.~\eqref{v3v4}
together with the adjacency matrices 
in Figs.~\ref{fig4} and~\ref{fig5} indicate 
that the no-crossing theorem breaks down in 
higher-dimensional systems. Furthermore, 
we observe that our crossings, both for the model 
problem as well as for the $2S$--$2S$ system, involve
situations where the couplings 
are indirect and the admixtures at the crossing point
are between levels which are displaced from each other in the 
adjacency graph by at least two 
elementary steps.
These observations could be of interest beyond the 
the concrete problem studied here, in the
context of a breakdown of the no-crossing theorem
in higher-dimensional quantum mechanical systems.
An improved understanding of the 
2S--2S interaction is important for progress 
in the 2S hyperfine measurement by optical methods,
using an atomic beam~\cite{FiKoKaHa2002,KoFiKaHa2004,KoEtAl2009}.

Attempts to study the hyperfine-resolved interaction
have been made, but no reference has been 
made to the resolution of the hyperfine structure~\cite{JoEtAl2002,SiKoHa2011}.
The current approach leads to a solution,
with partial results being 
presented in Eq.~\eqref{mat} 
and Fig.~\ref{fig7} and finer 
details being relegated to Ref.~\cite{JeEtAl2016vdWii}.


\section*{Acknowledgments}

The authors acknowledge support from the 
National Science Foundation (Grant PHY--1403973).

\end{document}